\documentclass[12pt]{JHEP3}

\usepackage{amscd,amsmath,amssymb,amsfonts,xspace,mathrsfs}
\usepackage{graphicx}
\usepackage{fancyhdr}
\usepackage{rotating}
\usepackage{psfrag}
\usepackage{epsfig}
\usepackage{url}
\usepackage{bbold}
\usepackage{undertilde}

\allowdisplaybreaks
\newcommand{\subsubsubsection}[1]{
\vspace{0.3cm}
\noindent {\it #1}
\vspace{0.2cm}
}

\newcounter{tabl}
\newcommand{\be}{\begin{equation}}
\newcommand{\ee}{\end{equation}}
\def\ba{\begin{eqnarray}}
\def\ea{\end{eqnarray}}
\newcommand{\beq}{\begin{eqnarray}}
\newcommand{\eeq}{\end{eqnarray}}
\newcommand{\bea}[2]{\be\label{#2}\begin{array}{#1}}
\newcommand{\eea}{\end{array}\ee}

\def\det{\,{\rm det}\, }
\def\diag{{\rm diag}}

\def\Im{\,{\rm Im}\, }

\def\({\left(}
\def\){\right)}
\def\[{\left[}
\def\]{\right]}
\def\p{\partial}
\newcommand{\de}{\mathrm{d}}
\newcommand{\I}{\mathrm{i}}

\def\11{1\!\! 1}

\def\hf{\frac{1}{2}}

\def\eps{\varepsilon}

   \def\CD {{\cal D}}
   \def\CE {{\cal E}}
   
   \def\CG {{\cal G}}
   \def\CH {{\cal H}}

   \def\CL {{\cal L}}
   \def\CM {{\cal M}}
   \def\CN {{\cal N}}

   \def\CS {{\cal S}}

   \def\CX {{\cal X}}

\newcommand{\tX}{\lefteqn{\smash{\mathop{\vphantom{<}}\limits^{\;\sim}}}X}
\newcommand{\tE}{\lefteqn{\smash{\mathop{\vphantom{<}}\limits^{\;\sim}}}E}

\newcommand{\tP}{\lefteqn{\smash{\mathop{\vphantom{<}}\limits^{\;\sim}}}P}

\newcommand{\Et}{\lefteqn{\smash{\mathop{\vphantom{\Bigl(}}\limits_{\sim}
\atop \ }}E}
\newcommand{\Pt}{\lefteqn{\smash{\mathop{\vphantom{\Bigl(}}\limits_{\sim}
\atop \ }}P}

\newcommand{\Nt}{\lefteqn{\smash{\mathop{\vphantom{\Bigl(}}\limits_{\sim}
\atop \ }}N}
\newcommand{\Mt}{\lefteqn{\smash{\mathop{\vphantom{\Bigl(}}\limits_{\,\sim}
\atop \ }}M}

\newcommand{\teps}{\tilde{\eps}}
\newcommand{\epst}{\lefteqn{\smash{\mathop{\vphantom{\Bigl(}}\limits_{\!\scriptstyle{\sim}}\atop \ }}\eps}

\def\hCG{\hat\CG}

\def\hCD{\hat\CD}
\def\hCH{\hat\CH}
\def\hN{\hat N}
\def\hM{\hat M}
\def\hNt{\hat\Nt}
\def\hMt{\hat\Mt}
\def\hn{\hat n}

\def\omp{\omega_+}
\def\omm{\omega_-}
\def\ompm{\omega_\pm}

\def\ompi#1{\omega_{+,#1}}
\def\ommi#1{\omega_{-,#1}}

\def\Xp{X_+}
\def\Xm{X_-}

\def\tXp{\tX_+}
\def\tXm{\tX_-}

\def\tXpi#1{\tX_{+,#1}}
\def\tXmi#1{\tX_{-,#1}}

\def\eep{e_+}
\def\eem{e_-}
\def\eepm{e_\pm}

\def\eepi#1{e_{+,#1}}
\def\eemi#1{e_{-,#1}}
\def\eepmi#1{e_{\pm,#1}}

\def\tPpi#1{\tP_{+,#1}}
\def\tPmi#1{\tP_{-,#1}}

\def\Ptpi#1{\Pt_{+,#1}}
\def\Ptmi#1{\Pt_{-,#1}}

\def\tEpi#1{\tE_{+,#1}}
\def\tEmi#1{\tE_{-,#1}}
\def\tEpmi#1{\tE_{\pm,#1}}

\def\chip{\chi_+}
\def\chim{\chi_-}

\def\chipi#1{\chi_{+,#1}}
\def\chimi#1{\chi_{-,#1}}

\def\etapi#1{\eta_{+,#1}}
\def\etami#1{\eta_{-,#1}}

\def\Np{N_+}
\def\Nm{N_-}
\def\Npm{N_\pm}

\def\Ntp{\Nt_+}
\def\Ntm{\Nt_-}
\def\Ntpm{\Nt_\pm}

\def\gp{g_+}
\def\gm{g_-}

\def\mett{{\mathbf g}}

\def\PhiG{\mathscr{G}}
\def\PhiD{\mathscr{D}}
\def\PhiS{\mathscr{V}}
\def\PhiH{\mathscr{H}}
\def\Scon{\CS}

\def\CGpi#1{\PhiG_{+,#1}}
\def\CGmi#1{\PhiG_{-,#1}}
\def\CGpmi#1{\PhiG_{\pm,#1}}

\def\CDpi#1{\PhiD_{+,#1}}
\def\CDmi#1{\PhiD_{-,#1}}
\def\CDpmi#1{\PhiD_{\pm,#1}}

\def\mhD{\hat\PhiD^{\rm int}}
\def\mHp{\PhiH_+^{\rm int}}
\def\mHm{\PhiH_-^{\rm int}}

\title{Canonical structure of Tetrad Bimetric Gravity}

\author{Sergei Alexandrov
\\

{\it Universit\'e Montpellier 2, Laboratoire Charles Coulomb, UMR 5221,\\ F-34095, Montpellier, France}
\\

{\tt e-mail:\email{salexand@univ-montp2.fr}}
}

\abstract{We perform the complete canonical analysis of the tetrad formulation of bimetric gravity and confirm that
it is ghost-free describing the seven degrees of freedom of a massless and a massive gravitons.
In particular, we find explicit expressions for secondary constraints, one of which is responsible for removing the ghost,
whereas the other ensures the equivalence with the metric formulation. Both of them have a remarkably
simple form and, being combined with conditions on Lagrange multipliers, can be written in a covariant way.}

\begin{document}

\section{Introduction}

The idea of considering two gravitational fields described by the Einstein-Hilbert actions and coupled
by a non-derivative interaction term goes back to the 1960's \cite{Salam:1969rq,Isham:1970gz}.
The corresponding theory is known as {\it bimetric gravity}. It is closely related to massive gravity
models which can be obtained from bimetric gravity by setting one of the gravitational fields to a background value.
Their linearization is captured by the Fierz-Pauli free theory \cite{Fierz:1939ix} describing 5 degrees of freedom of a massive graviton.
However, it was shown in \cite{Boulware:1973my} that generically at the non-linear level there is an additional propagating
scalar field with a wrong sign kinetic term, known as the Boulware-Deser ghost. This pathology plagues both
massive gravity and bigravity theories and obstructed the research in this domain for almost forty years.

A resolution of this problem was found only recently in the works of de Rham, Gabadadze and Tolley
\cite{deRham:2010ik,deRham:2010kj,deRham:2011rn} where they discovered 3 interaction potentials which
were shown to be ghost-free \cite{Hassan:2011ea,Mirbabayi:2011aa,Hassan:2012qv}.
These results were then extended to the case of two dynamical metrics \cite{Hassan:2011tf,Hassan:2011zd} which led to
a proposal for bimetric gravity models free of the ghost pathology.
These findings triggered a lot of works which studied various properties of these models,
see \cite{Hinterbichler:2011tt} and references therein.

However, a serious drawback of the proposed ghost-free potentials was their awkward form involving matrix square roots
of the product of the two metrics. This drawback was overcome in a beautiful paper \cite{Hinterbichler:2012cn}
where the complicated potentials of the metric formulation have been shown to acquire a very simple and natural form in
the tetrad formalism (see also \cite{Chamseddine:2011mu}).
Denoting the two sets of tetrad one-forms by $\eepm^I$, the general interaction term argued to lead to the ghost-freeness
was found to be
\be
S_{\rm int}
=-m^2\int \eps_{IJKL}\(\frac{\beta_1}{6}\, \eep^I\wedge \eem^J\wedge \eem^K\wedge \eem^L
+\frac{\beta_2}{4}\, \eep^I\wedge \eep^J\wedge \eem^K\wedge \eem^L
+\frac{\beta_3}{6}\, \eep^I\wedge \eep^J\wedge \eep^K\wedge \eem^L  \),
\label{intpot}
\ee
where $m^2$ is an overall mass parameter and $\beta$'s are coupling constants multiplying the three ghost-free potentials.
This simplification strongly indicates that the tetrads are
the right variables to formulate and study the massive gravity and bigravity models.
Furthermore, the tetrad formulation makes transparent the appearance of the primary constraint removing the Boulware-Deser ghost
because all interaction terms \eqref{intpot} are linear in the lapse and shift functions of both metrics.

Despite of all evidences and arguments given in \cite{Hassan:2011ea,Hinterbichler:2012cn} (see also \cite{Deffayet:2012nr,Comelli:2013txa}),
the ghost-freeness of the bigravity models, and of their reformulation in terms of tetrads in particular,
seems to be a controversial issue in the literature so far. Several groups argued that the ghost is not actually removed
\cite{Kluson:2013cy,Chamseddine:2013lid,Kluson:2013lza,Kluson:2013aca}. In particular, the last of the cited papers indicated
that the tetrad reformulation of bimetric gravity involves several new features, such as the presence of local Lorentz invariance,
which might affect its Hamiltonian formulation in a non-trivial way. Moreover, it proposed a canonical analysis of the tetrad bigravity
and concluded that the theory propagates 8 degrees of freedom including the scalar ghost.
However, this analysis was based on the decomposition of variables borrowed from \cite{Damour:2002ws} which spoiled the main
property of the potential \eqref{intpot} --- the linearity in the lapse functions.
As a result, it became much more cumbersome than it was expected to be and its final conclusions are not really trustworthy.

The aim of this paper is to settle down the issue of the ghost-freeness by performing the thorough canonical analysis
of bimetric gravity in the tetrad formulation. In contrast to \cite{Kluson:2013aca}, we take advantage of the linearity property
of the interaction term \eqref{intpot}, which immediately allows to get the complete set of primary constraints.
We compute the algebra of these constraints and find the secondary constraints resulting from
the stabilization procedure for the primary ones.
As a result, we obtain that the theory has only seven degrees of freedom so that the Boulware-Deser ghost is indeed absent.

We wish to emphasize that the secondary constraints are evaluated explicitly and found to be remarkably simple.
Furthermore, being combined with conditions on Lagrange multipliers, also resulting from the stabilization procedure,
they can be put in a covariant form. There are two such constraints. One of them reproduces the ``symmetricity"
condition on the tetrads which was shown to ensure the equivalence to the metric formulation of bigravity
\cite{Hinterbichler:2012cn,Deffayet:2012zc}. The second constraint is responsible for removing the ghost and,
in the covariant form, turns out to coincide with the field equation found in \cite{Deffayet:2012nr}.
Thus, our results are perfectly consistent with general expectations and previous findings at the Lagrangian level.

The paper is organized as follows.
In the next section we present the model to be analyzed.
In section \ref{sec-structure} we describe its $3+1$ decomposition, the stabilization procedure for all constraints, and the resulting
canonical structure of the theory. Although this requires the explicit knowledge of the constraint algebra,
we postpone its discussion to section \ref{sec-proofs}.
Section \ref{sec-concl} contains our conclusions. Appendix \ref{sec-HP} sets our conventions and
presents a review of the canonical analysis of the Hilbert-Palatini formulation of general relativity,
whereas appendix \ref{sec-useful} provides some relations used in calculations of the main body of the paper.

\section{Bimetric gravity in the tetrad formulation}
\label{sec-model}

The action of bimetric gravity that we consider in this paper is represented in the following form
\be
S[\eepm,\ompm]=S_{\rm HP}[\eep,\omp]+S_{\rm HP}[\eem,\omm]+S_{\rm int}[\eep,\eem],
\label{totalaction}
\ee
where the dynamics of each of the two sectors is described by\footnote{We absorbed the two Newton constants by rescaling
the tetrad fields $\eepm^I$ and the cosmological constants $\Lambda_\pm$ in the two sectors.
They can always be restored by dimensional analysis.}
\be
\begin{split}
S_{\rm HP}[e,\omega]=&\,\frac14\int \eps_{IJKL} e^I\wedge e^J \wedge \(F^{KL}(\omega)+\frac{\Lambda}{24}\, e^K\wedge e^L\),
\\
F(\omega)^{IJ}=&\, \de \omega^{IJ}+{\omega^I}_K \wedge \omega^{KJ},
\label{HPaction}
\end{split}
\ee
whereas the coupling between them is given by the interaction term $S_{\rm int}$ defined in \eqref{intpot}.
The action \eqref{HPaction} is the standard Hilbert--Palatini action of general relativity in the first order formulation
where the spin connection $\omega^{IJ}$ is considered as an independent dynamical field.
Upon solving its equations of motion
\be
D e^I\equiv \de e^I+{\omega^I}_J\wedge  e^J=0,
\label{Cartaneq}
\ee
the Hilbert-Palatini action reduces to the one of the tetrad formalism.
In this work we prefer to keep $\omega^{IJ}$ independent since this allows to write all constraints as polynomials
in dynamical fields. The price to pay for this will be the presence of additional second class constraints restricting
some of the components of the spin connection (see below).

In \cite{Hinterbichler:2012cn} it was shown that the interaction term \eqref{intpot} reproduces the ghost-free potentials
of massive gravity found in \cite{deRham:2010kj,Hassan:2011vm} in the metric formulation provided the tetrads satisfy
the following ``symmetricity" condition
\be
\eta^{[IK}\eepi{K}^\mu\eemi{\mu}^{J]}=0,
\label{symcond}
\ee
where $\eepmi{I}^\mu$ is the tetrad inverse to $\eepmi{\mu}^I$ and square brackets
denote the anti-symmetrization of the corresponding indices.
Moreover, this constraint was argued to follow
from equations of motion which implies that the metric and tetrad formulations should be dynamically equivalent.
Precise conditions when these statements become true were further analyzed in \cite{Deffayet:2012nr,Deffayet:2012zc}.
For our purposes, it will be convenient to note that \eqref{symcond} can be rewritten in the way
which does not involve inverse tetrads and reads as follows
\be
\eta_{IJ} \eepi{[\mu}^I \eemi{\nu]}^J=0.
\label{covconstr}
\ee

It is clear that, whereas the theory at $m^2=0$ possesses 20 gauge symmetries
(6 local Lorentz rotations and 4 diffeomorphisms in each of the two sectors),
in the interacting theory only the diagonal gauge transformations survive.
The off-diagonal symmetries are explicitly broken by the mass term.
Nevertheless, as we will see, they leave a trace in the theory generating second class constraints.

\section{Canonical structure of bimetric gravity}
\label{sec-structure}

\subsection{Preliminary comments}
\label{subsec-prelim}

This section is central in the paper. Here we present the results of the canonical analysis of the bimetric gravity
described by the action \eqref{totalaction}. To facilitate understanding of the resulting structure,
all details involving non-trivial calculations are postponed to section \ref{sec-proofs}.

It is worth to note that our analysis is very close to the one done in \cite{Alexandrov:2012yv}
where bimetric gravity in the so called chiral formulation has been considered.
It is described by essentially the same action as \eqref{totalaction}
where however the Lorentz gauge group is replaced by its (complexified) chiral subgroup $SU(2)$.\footnote{In fact, the starting point
in \cite{Alexandrov:2012yv} was chosen to be the so called Plebanski action \cite{Plebanski:1977zz} which is written in terms of two-forms
$B\sim e\wedge e$. However, it is completely equivalent to the action written in terms of tetrads so that this difference can be ignored.
Its only effect is that one had to restrict the interaction term \eqref{intpot} to the case of vanishing $\beta_1$ and $\beta_3$
since the corresponding potentials cannot be formulated in terms of two-forms $B$.}
In \cite{Alexandrov:2012yv} this chiral theory was put into the Hamiltonian formulation and was shown to be ghost-free.
Despite a close resemblance of the two theories and that many results turn out to be indeed very similar,
the difference in the gauge group does not allow to immediately apply the conclusions obtained in one case to another.
Besides, the chiral formulation suffers from necessity to impose reality conditions to extract a real theory, and
the problem of finding the reality conditions relevant for the massive case has not been solved in \cite{Alexandrov:2012yv}.
In fact, as we discuss in Conclusions, the results of this paper suggest a natural candidate for such reality conditions
upon using of which the two theories become identical.

\subsection{The phase space and primary constraints}
\label{subsec-decomp}

The first step to be done is the $3+1$  decomposition of the action \eqref{totalaction}. To this end, we
decompose the tetrad one-form (in both sectors) as follows\footnote{The tilde over or under variable indicates whether
it appears as a spatial density of positive or negative weight.}
\be
e^I=\(\Nt \tX^I+N^a e_a^I\)\de t + e_a^I\de x^a,
\label{dec-tetrad}
\ee
where $e_a^I$ and $\tX^I$ satisfy
\be
\eta_{IJ}\tX^Ie_a^J=0,
\qquad
\eta_{IJ}\tX^I \tX^J=-g,
\qquad
\eta_{IJ} e_a^I e_b^J=g_{ab},
\label{propvectors}
\ee
where $g_{ab}$ is the induced metric on the spatial slice and $g$ is its determinant.
If one wishes, one can solve these relations introducing the following explicit parametrization
\be
\begin{split}
\tX^I=&\,\sqrt{h}\(1,\chi^i\),
\\
e_a^I=&\,\(E_a^j\chi_j,E_a^i\),
\end{split}
\label{parameX}
\ee
where the index $I$ runs over $\{0,i\},$ $i=1,2,3$, and $\sqrt{h}=\det E_a^i$.

The decomposition of the Hilbert-Palatini action \eqref{HPaction} and its canonical structure are recalled in appendix \ref{sec-HP}.
There are two equivalent ways to formulate them. In one way, which preserves the explicit Lorentz covariance, the phase space
is parametrized by
\be
\omega_a^{IJ}\quad  {\rm and} \quad
\tP^a_{IJ}\equiv \frac14\, \teps^{abc} \eps_{IJKL} e_b^K e_c^L.
\label{canvar}
\ee
However, the variables $\tP^a_{IJ}$ are not all independent and therefore satisfy certain constraints (see \eqref{conphi}) which
in turn generate secondary constraints on $\omega_a^{IJ}$ (see \eqref{conpsi}).
Altogether they are second class and affect the symplectic structure
which is now given by Dirac brackets. In particular, the commutator between the variables \eqref{canvar} is not trivial anymore
and can be found in \eqref{commomP}.

The second possibility is to use the parametrization \eqref{parameX}.
It allows to explicitly solve the above second class constraints. Using the variables introduced in
\cite{Alexandrov:1998cu}, the phase space is then parametrized by two canonical pairs
\be
(\eta_a^i,\tE^a_i)\qquad {\rm and }\qquad (\omega^i, \chi_i).
\label{noncovvar}
\ee
Here $E^a_i$ is the inverse to $E_a^i$, $\tE^a_i$ denotes its densitized version,
and $\eta_a^i$, $\omega^i$ are related to the spatial components of the spin connection as
\be
\omega_a^{0i}=\eta_a^i-\omega_a^{ij}\chi_j,
\qquad
\omega_a^{ij}=\eps^{ijk}r_{kl}\Et_a^l +\Et_a^{[i}\omega^{j]},
\label{solom}
\ee
where $r_{ij}$ is symmetric and fixed by equations \eqref{solr} representing the solution of the secondary second class constraints.
Using these relations and the canonical Poisson brackets for the variables \eqref{noncovvar},
one can check that the covariant variables \eqref{canvar} satisfy the same commutation relations \eqref{commomP} as
in the covariant approach.

Substituting the decomposed Hilbert-Palatini action and the interaction term into \eqref{totalaction}, and changing the
variables playing the role of Lagrange multipliers to extract the symmetric and anti-symmetric combinations of primary constraints,
one finds the following expression for the total action
\be
\begin{split}
S=&\,\int \de t\int\de^3 x\Bigl[\tEpi{i}^a\p_t \etapi{a}^{i}+\tEmi{i}^a\p_t \etami{a}^{i}+\chipi{i}\p_t\omp^i+\chimi{i}\p_t\omm^i
\Bigr.
\\
&\, \Bigl.\qquad
+n^{IJ}\CG_{IJ}+\hn^{IJ}\hCG_{IJ}+N^a\CD_a+\hN^a\hCD_a+\Nt\CH+\hNt\hCH\Bigr].
\end{split}
\label{decomtotact}
\ee
In the absence of the mass term, $\CG_{IJ}, \CD_a, \CH$ and $\hCG_{IJ}, \hCD_a, \hCH$ would be first class constraints
generating diagonal and off-diagonal transformations, respectively, corresponding to local Lorentz rotations, spatial and time diffeomorphisms.
However, as one can see from \eqref{doffd-constr}, the mass term affects $\hCD_a$, $\CH$ and $\hCH$ and changes the constraint algebra.
It is explicitly calculated below in section \ref{sec-proofs}. As a result, $\CG_{IJ}$ and $\CD_a$ continue to weakly commute
with all other constraints and therefore remain first class generating diagonal Lorentz and spatial diffeomorphism transformations.
On the other hand, the commutators of other constraints acquire non-vanishing contributions and require a careful study of
the stability of the corresponding constraints under time evolution.

\subsection{Stabilization procedure}
\label{subsec-stabil}

Let us present the results of the analysis of stability of the remaining primary constraints one by one.
Some missing details can be found in section \ref{sec-proofs}.

\subsubsubsection{Stability of $\hCG_{IJ}$}

Preservation of $\hCG_{IJ}$ under time evolution generates 6 equations.
Due to a very special property of the constraint algebra (see \eqref{prop-commCGH} below),
they can be split into 3 secondary constraints
\beq
\Scon^a&=&\teps^{abc} \eta_{IJ} \eepi{b}^I \eemi{c}^J\approx 0
\label{constrCcov}
\eeq
and 3 conditions on the Lagrange multipliers which fix the variables $\hN^a$
\be
\begin{split}
\hN^a=&\,
-\hf\, \mett^{ab}\,\Bigl[ \Ntp\(\eta_{MN}\tXp^M e_{-b}^N\)-\Ntm\(\eta_{MN}\tXm^M e_{+b}^N\)\Bigr],
\end{split}
\label{solhN}
\ee
where we introduced the metric defined by the tetrads from the two sectors
\be
\mett_{ab}=\eta_{IJ} \eepi{a}^I \eemi{b}^J
\label{meth}
\ee
and assumed its invertibility. The last condition is equivalent to $\det\mett_{ab}=\eta_{IJ}\Xp^I\Xm^J\ne 0$ what we assume
to be the case throughout the paper.
Note that the physical meaning of the secondary constraint \eqref{constrCcov} is that the metric $\mett_{ab}$
should be symmetric.
Furthermore, taking into account that $\hN^a=\hf\, (\Np^a-\Nm^a)$, it is easy to check that
altogether the 6 equations \eqref{constrCcov} and \eqref{solhN} represent the $3+1$ decomposition
of the covariant ``symmetricity" condition \eqref{covconstr}. Thus, at the canonical level this condition,
and as a result the equivalence of the tetrad
formulation to the metric one, follow from stabilization of the off-diagonal Gauss constraint.

\subsubsubsection{Stability of $\hCD_a$ and $\Scon^a$}

It is convenient to consider together these two constraints because they both have
non-vanishing commutators with the off-diagonal Gauss constraint $\hCG_{IJ}$. As a result, their stability conditions
produce 6 equations which fix the Lagrange multipliers $\hn^{IJ}$. Again this requires the invertibility of a matrix
which can be represented in terms of two $3\times 6$ blocks $(\CM^a_{IJ}, \CN^a_{IJ})$ defined by
\be
\begin{split}
\CM^a_{IJ}\equiv&\,
\eps_{IJKL}\teps^{abc}\(\beta_1 \eemi{b}^K \eemi{c}^L+2\beta_2 \eepi{b}^K\eemi{c}^L+\beta_3 \eepi{b}^K\eepi{c}^L\),
\\
\CN^a_{IJ}\equiv& \, \eta_{IK}\eta_{JL}\teps^{abc}\eepi{b}^K\eemi{c}^L.
\end{split}
\label{defM}
\ee
Near the symmetric background $\eepi{a}^I\approx \eemi{a}^I$ this is equivalent to a simple condition on the parameters
of the mass term
\be
\beta_1+2\beta_2+\beta_3 \ne 0
\ee
which indeed holds if we require, for example, the existence of a flat space solution \cite{Hinterbichler:2012cn}.
As a result, we see that the three sets of constraints $\hCG_{IJ}$, $\hCD_a$ and $\Scon^a$ are conjugated to each other
and appear to be second class.

\subsubsubsection{Stability of $\CH$ and $\hCH$}

Once we take into account solutions for Lagrange multipliers $\hN^a$ and $\hn^{IJ}$ obtained at the previous steps,
it turns out the stability of $\CH$ and $\hCH$ reduces to one and the same condition giving rise to a secondary constraint.
It is given by the following expression
\be
\begin{split}
\Psi=&\, \{\CH,\hCH\}
+\[ \(\tPpi{IJ}^a-\tPmi{IJ}^a\)\{\hCD_a,\CH\}
-\(\tPpi{IJ}^a+\tPmi{IJ}^a\)\{\hCD_a,\hCH\}\]\frac{\tXp^I \tXm^J}{(\tXp\tXm)\vphantom{\tilde X}}
\\
&\, -2\tPpi{IJ}^a \{\hCD_a,\hCD_b\}\tPmi{KL}^b\,\frac{\tXp^I \tXm^J \, \tXp^K \tXm^L}{\(\tXp\tXm\)^2}\approx 0,
\end{split}
\label{genPsi}
\ee
where we used the formula \eqref{inversemet} for the inverse metric $\mett^{ab}$
and dropped the $\delta$-function factor in the commutators.
A straightforward calculation, based on the results for the constraint algebra given in \eqref{algebra} below and various properties
presented in appendix \ref{sec-useful}, then leads to an extremely simple result
\be
\Psi= -\frac{m^2\gp\gm}{2(\tXp \tXm)\vphantom{\tilde X}}\,\CM^a_{IJ} \(\ompi{a}^{IJ}-\ommi{a}^{IJ}\).
\label{Psiexpr}
\ee

Furthermore, as it happened with the constraint \eqref{constrCcov}, $\Psi$ can be written in a covariant form being
combined with equations on the Lagrange multipliers. Namely, it is easy to see that it appears as the zeroth component of
the following covariant vector
\be
\CE^\mu=\teps^{\mu\nu\rho\sigma}\eps_{IJKL}\(\beta_1 \eemi{\nu}^K \eemi{\rho}^L+2\beta_2 \eepi{\nu}^K \eemi{\rho}^L
+\beta_3 \eepi{\nu}^K \eepi{\rho}^L\)\(\ompi{\sigma}^{IJ}-\ommi{\sigma}^{IJ}\).
\label{vecPsi}
\ee
The remaining components $\CE^a$ can be checked to coincide with a linear combination of the stability condition for $\hCD_a$,
fixing some components of $\hn^{IJ}$, and other constraints.\footnote{To this end, one should take into account
the relation of the Lagrange multipliers $\hn^{IJ}$ to the spin connection following from \eqref{nom} and \eqref{newLm}
$$
\hn^{IJ}=\hf\[\(\ompi{0}^{IJ}-\Np^a \ompi{a}^{IJ}\)-\(\ommi{0}^{IJ}-\Nm^a \ommi{a}^{IJ}\)\].
\label{omegahn}
$$
}
This implies that one has a vector equation $\CE^\mu\approx 0$. Remarkably, it seems to reproduce eq. (70) in \cite{Deffayet:2012nr}
found in the context where one of the tetrads is kept fixed. Here we see that in bimetric gravity, from the canonical point of view,
one of these equations appears as the secondary constraint ensuring the stability of the two primary Hamiltonian constraints.

\subsubsubsection{Stability of $\Psi$}

Finally, it remains to consider the stability condition for the secondary constraint $\Psi$ \eqref{Psiexpr}.
Generically, it has non-vanishing commutators with the Hamiltonian constraints $\CH$ and $\hCH$.
Therefore, its stability condition having the form
\be
\{\Psi, \CH(\Nt)\}+\{\Psi,\hCH(\hNt)\}+\cdots=0
\label{stabPsi}
\ee
is expected to fix one of the Lagrange multipliers $\Nt$ or $\hNt$ in terms of the other.
For this to be true, the equation \eqref{stabPsi} should be algebraic in these variables,
i.e. it should not contain spatial derivatives acting on them.
An elegant method to prove this just using the Jacobi identity and the fact that a similar property
is true for the commutator \eqref{genPsi} defining $\Psi$ itself, has been developed in \cite{Alexandrov:2012yv}.
Since it applies directly to our case as well, we do not repeat it here, but refer the interested reader to section 4.4 of that paper.
Thus, the secondary constraint forms a second class pair with one of the Hamiltonian constraints.

\subsection{Summary}
\label{subsec-sum}

Let us summarize the resulting canonical structure.
Once we use the parametrization \eqref{noncovvar} of the phase space of the Hilbert-Palatini formulation,
in which the second class constraints of that formulation are explicitly resolved, the phase space
of tetrad bimetric gravity is $2\times 24=48$ dimensional. It caries 10 first class constraints $\CG_{IJ}$, $\CD_a$
and\footnote{Note that it is wrong to expect that one of the Hamiltonian constraints, for example, the diagonal one $\CH$,
will be first class by itself and generate diagonal time diffeomorphisms. It is well known that
such transformations are generated by the total Hamiltonian which is by construction first class for
diffeomorphism invariant systems. In our case, $H_{\rm tot}$ can be obtained as a linear combination of primary constraints
with the coefficients given by the Lagrange multipliers expressed in terms of their solutions of the stability conditions.
Once all such conditions are solved, all Lagrange multipliers of constraints which are second class
become proportional to $\Nt$ and thus it is sufficient to extract the expression it multiplies.\label{footH}}
$H_{\rm tot}$ generating diagonal gauge transformations and 14 second class constraints
$\hCG_{IJ}$, $\hCD_a$, $\Scon^a$, $\Psi$ and either $\CH$ or $\hCH$.
Thus, we remain with $48-2\times 10-14=14$ dimensional space describing 7 degrees of freedom,
which clearly can be identified with those of one massless and one massive graviton. The scalar ghost degree of freedom
is absent.

\section{Constraint calculus}
\label{sec-proofs}

In this section we compute the constraints and their algebra and show how they
lead to the results presented in the previous section.

First of all, let us perform the $3+1$ decomposition of the interaction term \eqref{intpot} of the bigravity action.
Substituting \eqref{dec-tetrad} and using some of the relations \eqref{usefulrel}, one arrives at
\be
S_{\rm int}=\int d^3 x\, \[\hf\,(\Np^a-\Nm^a)\mhD_a +\Ntp\mHp+\Ntm\mHm\],
\ee
where
\beq
\mhD_a
&=& 2 m^2\Bigl(\beta_1\eta_{IJ} \eepi{a}^I\tXm^J-\beta_2 \eps^{IJKL}\epst_{abc}\tPpi{IJ}^b\tPmi{KL}^c-\beta_3\eta_{IJ} \eemi{a}^I\tXp^J\Bigr) ,
\nonumber\\
\mHp
&=& m^2 \Bigl(\beta_1\eta_{IJ} \tXp^I\tXm^J-2\beta_2 \gp\Ptpi{a}^{IJ}\tPmi{IJ}^a -\beta_3 \gp \gp^{ab} \eepi{a}^I \eemi{b}^J\eta_{IJ}\Bigr),
\label{massH}
\\
\mHm
&=& m^2 \Bigl(-\beta_1 \gm \gm^{ab} \eemi{a}^I \eepi{b}^J\eta_{IJ}-2\beta_2 \gm\Ptmi{a}^{IJ}\tPpi{IJ}^a +\beta_3 \eta_{IJ} \tXp^I\tXm^J\Bigr).
\nonumber
\eeq
Combining this result with decompositions of the two Hilbert-Palatini actions dependent of $+$ and $-$ variables,
and introducing symmetric and anti-symmetric combinations of the Lagrange multipliers
\be
n_\pm^{IJ}=n^{IJ}\pm \hn^{IJ},
\qquad
\Npm^a= N^a\pm \hN^a,
\qquad
\Ntpm= \Nt\pm \hNt,
\label{newLm}
\ee
one finds the total action \eqref{decomtotact} given above with the primary constraints
given by the following combinations
\be
\begin{array}{rclcrcl}
\CG_{IJ}&=&\CGpi{IJ}+\CGmi{IJ},
&\qquad&
\hCG_{IJ}&=&\CGpi{IJ}-\CGmi{IJ},
\\
\CD_{a}&=&\CDpi{a}+\CDmi{a},
&\qquad&
\hCD_a&=&\CDpi{a}-\CDmi{a}+\mhD_a,
\\
\CH&=&\PhiH_+ +\PhiH_-+\mHp+\mHm,
&\qquad&
\hCH&=&\PhiH_+ -\PhiH_- +\mHp-\mHm.
\end{array}
\label{doffd-constr}
\ee
The explicit expressions for $\CGpmi{IJ},\CDpmi{a},\PhiH_\pm$ can be found in \eqref{HPconstr} and \eqref{defD}
where all variables should appear with $+$ or $-$ subscript, respectively.

The main non-trivial calculation needed for this work is evaluation of the algebra of the constraints \eqref{doffd-constr}.
Although it is a tedious exercise, it can be done quite efficiently using the action of the constraints on the basic variables
given in \eqref{constract}. Moreover, what we need is only the contributions non-vanishing on the constraint surface.
Restricting the attention to such contributions only, one finds\footnote{Here all repeated indices are assumed
to be contracted with the Minkowski metric $\eta_{IJ}$.}
\be
\begin{split}
\{\hCG(\hn),\hCD(\vec \hN)\}\approx &\,
m^2\int \de^3 x\,\hn^{IJ}\hN^a \mett_{ab} \CM^b_{IJ},
\\
\{\hCG(\hn),\CH(\Nt)\}\approx &\,
\frac{m^2}{2}\int \de^3 x\, \Nt \hn^{IJ}
\( \(\tXp e_{-a}\)-\(\tXm e_{+a}\)\)\CM^a_{IJ},
\\
\{\hCG(\hn),\hCH(\hNt)\}\approx &\,
\frac{m^2}{2}\int \de^3 x\, \hNt \hn^{IJ}
\( \(\tXp e_{-a}\)+\(\tXm e_{+a}\)\)\CM^a_{IJ},
\\
\{\hCD(\vec \hN),\hCD(\vec \hM)\}\approx &\,
-8m^2\int \de^3 x\, \hN^{[a} \hM^{b]}\Bigl[\beta_1 \tXmi{I} {\ompi{a}^{IJ}}\eepi{b}^J+\beta_3 \tXpi{I} {\ommi{a}^{IJ}}\eemi{b}^J
\Bigr.
\\
& \Bigl. \qquad
+\beta_2\epst_{abc}\eps^{IJKL}\(\tPpi{IJ}^c\ommi{d}^{KM}\tPmi{ML}^d+\tPmi{IJ}^c\ompi{d}^{KM}\tPpi{ML}^d\)\Bigr],
\\
\{\CH(\Nt),\hCD(\vec \hN)\}\approx &\,
2 m^2\int \de^3 x\, \Nt \hN^a\(\Delta_a^+ +\Delta_a^-\),
\\
\{\hCH(\hNt),\hCD(\vec \hN)\}\approx &\,
2 m^2 \int \de^3 x\,\hNt \hN^a\(\Delta_a^+ -\Delta_a^-\),
\\
\{\CH(\Nt),\hCH(\hMt)\}\approx&\,
4 m^2\int \de^3 x\, \Nt\hMt\Bigl[\beta_1 \tXpi{I}\tPmi{JK}^a \tXm^K
+\beta_2 \gp\gm\eps_{IJKL}\teps^{abc}\Ptpi{b}^{KM}\Ptmi{c}^{ML}
\Bigr.
\\
& \Bigl. \qquad
+\beta_3 \tXmi{I}\tPpi{JK}^a \tXp^K \Bigr]\(\ompi{a}^{IJ}-\ommi{a}^{IJ}\),
\end{split}
\label{algebra}
\ee
where $\CM^a_{IJ}$ was defined in \eqref{defM} and
\be
\begin{split}
\Delta_a^+ =&\,-\beta_1 \ompi{a}^{IJ} \tXpi{I} \tXmi{J}
+4\beta_2 \gp\tPmi{IK}^b\(\Ptpi{b}^{KJ}\ompi{a}^{IJ}-\Ptpi{a}^{KJ}\(\ompi{b}^{IJ}-\ommi{b}^{IJ}\)\)
\\
&\,
+\beta_3 \gp \gp^{bc} \eepi{c}^I\(  \ommi{a}^{IJ} \eemi{b}^J+\(\ompi{b}^{IJ}-\ommi{b}^{IJ}\)\eemi{a}^J\),
\\
\Delta_a^- =&\, \beta_1 \gm \gm^{bc} \eemi{c}^I\(  -\ompi{a}^{IJ} \eepi{b}^J+\(\ompi{b}^{IJ}-\ommi{b}^{IJ}\)\eepi{a}^J\)
\\
&\,
-4\beta_2 \gm\tPpi{IK}^b\(\Ptmi{b}^{KJ}\ommi{a}^{IJ}+\Ptmi{a}^{KJ}\(\ompi{b}^{IJ}-\ommi{b}^{IJ}\)\)
-\beta_3 \ommi{a}^{IJ} \tXpi{I} \tXmi{J}.
\end{split}
\label{defDelta}
\ee

Several comments are in order:
\begin{itemize}
\item
The constraints $\CG_{IJ}$ and $\CD_a$ generating diagonal Lorentz rotations and spatial diffeomorphisms commute weakly with all constraints,
and therefore appear to be first class,
consistently with the fact that these transformations remain gauge symmetries of the massive theory.
Note however that this is not true for the symmetric combination $\PhiS_{+,a}+\PhiS_{-,a}$ of the constraints appearing in \eqref{HPconstr}.
This justifies why we prefer to work in terms of $\CDpmi{a}$, which was proved to be convenient already in the case of pure general relativity.
Furthermore, as is clear from our results, the symmetric combination of the Hamiltonian constraints is not commuting either.
This is not surprising since the generator of diagonal time diffeomorphisms
should coincide with the total Hamiltonian (see footnote \ref{footH}).

\item
The result \eqref{algebra} is obtained on the surface of not only the primary constraints \eqref{doffd-constr}, but also
of the second class constraints $\psi_\pm^{ab}$ \eqref{conpsi} of the Hilbert-Palatini formulation
and the constraint $\Scon^a$ \eqref{constrCcov} introduced in the previous section. In particular, there are terms proportional to
$\Scon^a$ in the first three commutators.
Taking these into account, one arrives at the following crucial property
\be
\begin{split}
\{\hCG_{IJ},\CH(\Nt)\}=&\,
\hf\, \Nt\bigl( \(\tXp e_{-a}\)-\(\tXm e_{+a}\)\bigr)\,\mett^{ab}\{\hCG_{IJ},\hCD_b\} +O(\Scon^a),
\\
\{\hCG_{IJ},\hCH(\hNt)\}=&\,
\hf\,\hNt\bigl( \(\tXp e_{-a}\)+\(\tXm e_{+a}\)\bigr)\,\mett^{ab}\{\hCG_{IJ},\hCD_b\}   +O(\Scon^a).
\end{split}
\label{prop-commCGH}
\ee
It is this property which ensures that the stability condition for $\hCG_{IJ}$ fixes the Lagrange multipliers $\hN^a$ as in \eqref{solhN}
and generates $\Scon^a$ as secondary constraints.

\item
Another important property of the constraint algebra is the vanishing of
$\{\CH(\Nt),\CH(\Mt)\}$ and $\{\hCH(\hNt),\hCH(\hMt)\}$.
It makes easy to see that both stability conditions for $\CH$ and $\hCH$
generate the same secondary constraint \eqref{genPsi}, as we argued in the previous section.
However, the vanishing of these two commutators is not strictly speaking necessary for this result and,
for example, it fails in the formulation of \cite{Kluson:2013aca}.
But if they are non-vanishing, the commutator of two off-diagonal diffeomorphisms
acquires an additional contribution containing derivatives of the smearing functions, which should cancel
the above two commutators in the stability conditions. As a result, the non-trivial fact that two stability conditions
generate the same constraint is always expected to be true.

\end{itemize}

All the remaining facts mentioned in the previous section like, for example, the derivation of \eqref{Psiexpr}
or the check of $\CE^a\approx 0$, require straightforward, although sometimes lengthy calculations
and we refrain from showing them explicitly.

\section{Conclusions}
\label{sec-concl}

In this paper we have proven that the tetrad formulation of bimetric gravity,
with {all 3 possible mass terms included, propagates 7 degrees of freedom and therefore is ghost-free.
To this end, we developed its Hamiltonian formulation identifying all constraints and their nature.
In particular, we computed two secondary constraints which are both simple,
can be covariantized by combining them with conditions on Lagrange multipliers, and have a clear physical meaning:
one of them ensures the equivalence with the metric formulation and the other removes the Boulware-Deser ghost.

Although some of our calculations are relatively lengthy, they are certainly much easier than those in the metric formulation.
Moreover, the use of the first order formalism where the spin connection is considered as an independent variable
allowed to simplify them even comparing to the usual tetrad formulation.
For instance, without keeping the spin connection but expressing it in terms of derivatives of the tetrad,
it would be difficult to recognize that the secondary constraint \eqref{Psiexpr} has such a simple form.
Thus, keeping in mind that the reformulation of massive gravity in terms of tetrads has already proven to be extremely useful,
see for instance \cite{Deffayet:2012nr,Gabadadze:2013ria,Ondo:2013wka,Tamanini:2013xia,deRham:2013awa}, we would like to argue that
its first order version may be even more helpful for solving various problems.

Our analysis shows that the linearity of the interaction potential in the lapse and shift variables of the two metrics is not
sufficient for the absence of the ghost. One needs two additional ``miraculous" properties of the constraint algebra.
The first one is encoded in equations \eqref{prop-commCGH} and means that for those part of the off-diagonal Lorentz
transformations, which commute with $\hCD_a$, the commutators with the two Hamiltonian constraints should be proportional to each other.
If this was not the case, one would not generate the secondary constraints $\Scon^a$, but rather a condition on the Lagrange multipliers.
The second non-trivial property is that the stability of diagonal and off-diagonal Hamiltonian constraints
is ensured by the same secondary constraint $\Psi$.
In our case both these properties are intimately correlated with the linearity of the potential due to the Lorentz invariance.
If however one drops it and looks for more general ghost-free potentials as in \cite{Comelli:2013txa},
such properties might provide non-trivial conditions on the new potentials.

Of course, the ghost pathology is not the only problem experienced by a generic massive gravity model.
For instance, there are strong arguments that the ghost-free massive gravity suffers from superluminality in
the decoupling limit \cite{Gruzinov:2011sq,deFromont:2013iwa,Deser:2013eua}, contains tachyonic modes \cite{Deser:2013uy},
and exhibits ghost instabilities around cosmological homogeneous solutions
\cite{DeFelice:2012mx,DeFelice:2013bxa} (see however \cite{DeFelice:2013tsa}).
Some of these problems could be resolved if there exists a special choice of parameters corresponding
to a partially massless case where an additional gauge symmetry arises and
reduces the number of degrees of freedom of the massive graviton from 5 to 4
\cite{Deser:1983tm,Deser:1983mm}.
Although it was suggested that such a choice does exists and is moreover unique \cite{deRham:2012kf,Hassan:2012gz},
later it was argued that there are serious obstructions for that \cite{Deser:2013uy,deRham:2013wv,Deser:2013gpa}.
It would be interesting to see whether the formalism presented here can help in solving this problem.

There are actually several places in our analysis which under certain specific conditions
can lead potentially to a different phase space structure. Namely, we imposed the invertibility conditions on a few objects.
If in contrast they turn out to be non-invertible, which typically happens only in some degenerate situations,
the analysis may change. However, among them there is only one place where the cosmological constant, crucial for the partially massless
mechanism, plays a role. This is the stability condition for the secondary constraint $\Psi$ \eqref{stabPsi}.
We assumed that at least one of the commutators of $\Psi$ with the Hamiltonian constraints is non-vanishing.
If however they both vanish, this condition cannot be used to fix one of the Lagrange multipliers and
there is a chance that both $\CH$ and $\hCH$, properly adjusted by other constraints, are first class giving rise to a new gauge symmetry.
In \cite{Alexandrov:2012yv} such a possibility was investigated in a closely related chiral model and it was found that
there are solutions to all constraints for which the two commutators appearing in \eqref{stabPsi} are vanishing.
However, the geometric meaning of such solutions remain unclear.

Finally, we comment on the issue of reality conditions in the chiral model of \cite{Alexandrov:2012yv}.
As was mentioned in section \ref{subsec-prelim}, such conditions are needed to extract a real section of the model and
the ones relevant for the massive case have not been identified so far. The problem was that the natural conditions
demanding the reality of the complex triad fields $\tEpmi{i}^a$ of the chiral formulation lead to a trivial theory of two massless gravitons.
However, one should take into account that, even in the chiral formulation of general relativity,
such reality conditions can be achieved only by a special gauge choice. On the other hand, in the massive case one
of the two Lorentz gauge symmetries is broken and therefore such a simultaneous gauge choice for the two complex triads is not accessible.
Instead, one should impose the conditions which require the reality of only the spatial metric and its time evolution
\be
\Im \(\tEpmi{i}^a\tEpmi{i}^b\)=0,
\qquad
\p_t\Im\(\tEpmi{i}^a\tEpmi{i}^b\)=0.
\label{realcond}
\ee
These conditions can be identified \cite{Alexandrov:2005ng} with the second class constraints
of the Hilbert-Palatini formulation, \eqref{conphi} and \eqref{conpsi}, which in turn suggests that
the complex triad is related to the fields appearing in this work as follows
$\tE^a_i=2\tP^a_{0i}+\I{\eps_i}^{jk} \tP^a_{jk}$. (The complex field from the chiral model on the left
should not be confused with the real triad from \eqref{parameX}: the latter is the real part of the former.)
Given also that the difference between the actions of the chiral and our models vanishes on the surface of
the ``symmetricity" constraint $\Scon^a$, it is natural to conjecture that,
upon imposing the reality conditions \eqref{realcond}, the chiral model
\cite{Alexandrov:2012yv} reproduces exactly the tetrad bimetric gravity studied in this paper.\footnote{Note
that the model of \cite{Alexandrov:2012yv} is based on the so called chiral Plebanski action. There exists
also its non-chiral version which does not require any reality conditions. Remarkably, its modification
proposed in \cite{Alexandrov:2008fs} has been interpreted as a bimetric gravity \cite{Speziale:2010cf}.
However, it is plagued with the same scalar ghost as generic bimetric gravity models, and an interesting open problem is whether there exist
modifications of the non-chiral Plebanski action which lead to a ghost-free theory and
can be considered as analogues of the three potentials \eqref{intpot}.}

\section*{Acknowledgements}
The author is grateful to Josef Kluso\v{n} for valuable discussions.
This work is supported by contract ANR-09-BLAN-0041.

\appendix

\section{Hamiltonian formulation of the Hilbert-Palatini action}
\label{sec-HP}

In this appendix we present a brief review of the Hamiltonian formulation of the Hilbert-Palatini action
\be
S_{\rm HP}[e,\omega]=\frac14\int \eps_{IJKL} e^I\wedge e^J \wedge \(F^{KL}(\omega)+\frac{\Lambda}{24}\, e^K\wedge e^L\)
\label{HPact}
\ee
describing general relativity in the first order formalism.
It can be found, for instance, in the nice review \cite{Peldan:1993hi}
(for the original treatment, see \cite{Arnowitt:1960es,Deser:1976ay}),
however our exposition will be closer to
\cite{Alexandrov:1998cu,Alexandrov:2000jw} which use the same variables that we employ here.

First, let us fix our conventions. The tangent space indices $I,J=0,\dots,3$ are raised and lowered by means of the flat Minkowski metric
$\eta_{IJ}=\diag(-,+,+,+)$. The Levi-Civita symbol with flat indices is normalized as $\eps_{0123}=1$. On the other hand,
for the antisymmetric tensor density with spacetime indices we use $\teps^{0abc}=\teps^{abc}$. The symmetrization and anti-symmetrization
of indices denoted by $\{\cdot\, \cdot\}$ and $[\cdot\,\cdot]$, respectively, are both taken with weight 1/2.

The $3+1$ decomposition of the action \eqref{HPact} reads
\be
S_{\rm HP}=\int \de t\int\de^3 x\Bigl[\tP^a_{IJ}\p_t \omega_a^{IJ}+\omega_0^{IJ}\PhiG_{IJ}+N^a\PhiS_a+\Nt\PhiH\Bigr],
\ee
where
\beq
\PhiG_{IJ} &=&\p_a \tP^a_{IJ}+{\omega_{a,I}}^K\tP^a_{KJ}-{\omega_{a,J}}^K\tP^a_{KI}\equiv D_a\tP^a_{IJ},
\nonumber
\\
\PhiS_a &=& -\tP^b_{IJ} F_{ab}^{IJ},
\label{HPconstr}
\\
\PhiH &=& 2\tP^a_{IK}{\tP^{b,K}}_J \(F_{ab}^{IJ}-\frac{\Lambda}{12}\,\eps^{IJMN}\epst_{abc}\tP^c_{MN} \),
\nonumber
\eeq
and the momentum canonically conjugated to the spin connection was defined in \eqref{canvar}.
$\PhiG_{IJ}$, $\PhiS_a$ and $\PhiH$ are the primary constraints generating the gauge transformations of the theory.
It is convenient however to redefine one them as follows
\be
\PhiD_a \equiv\PhiS_a+\omega_a^{IJ}\PhiG_{IJ}=  \p_b\(\omega_a^{IJ}\tP^b_{IJ}\)-\tP^b_{IJ}\p_a\omega_b^{IJ},
\label{defD}
\ee
which can be achieved by the following redefinition of the Lagrange multiplier
\be
\omega_0^{IJ}=n^{IJ}+N^a\omega_a^{IJ}.
\label{nom}
\ee
The new constraint \eqref{defD} is more convenient because it is precisely the generator of spatial diffeomorphisms.

\subsection{Second class constraints and symplectic structure}
\label{subsec-sympl}

It turns out that, besides \eqref{HPconstr}, there are additional constraints to be imposed on the phase space parametrized by
$\omega_a^{IJ}$ and $\tP^a_{IJ}$. They stem from the fact that the 18 components of $\tP^a_{IJ}$ are defined in terms of only
12 independent components of $e_a^I$ and can be written as
\be
\phi^{ab}=\hf\,\eps^{IJKL}\tP^a_{IJ}\tP^b_{KL}=0.
\label{conphi}
\ee
Commuting them with the Hamiltonian, one generates secondary constraints
\be
\psi^{ab}=4\eps^{IJKL}\tP^{\{ a}_{IN}{\tP^{c,N}}_J D_c\tP^{b\}}_{KL}=0.
\label{conpsi}
\ee
It is easy to understand their physical meaning: they are nothing else but some components of
the first Cartan's structure equation \eqref{Cartaneq}, and thus allow to express 6 components
of the spin connection in terms of other fields. Most importantly is that, together with \eqref{conphi},
they form a second class pair and change the symplectic structure. The resulting Dirac brackets have been calculated in
\cite{Alexandrov:2000jw}. The fields $\tP^a_{IJ}$ remain commuting, their commutator with the spin connection is given by
\be
\{\omega_a^{IJ}(x), \tP^b_{KL}(y)\} =
\[\delta_a^b\delta^{IJ}_{KL}
+\frac{1}{4}\, \eps^{IJMN}\eps_{KLPQ}\(\tP^b_{MN}\Pt_a^{PQ}+\delta_a^b \tP^c_{MN}\Pt_c^{PQ}\)
\]\delta(x,y),
\label{commomP}
\ee
where we used another convenient notation
\be
\Pt_a^{IJ}=-g^{-1}g_{ab}\,\tP^{a,IJ}=g^{-1}\tX^{[I}e_a^{J]},
\label{defPt}
\ee
whereas the commutator of two spin connections is very complicated and will not be needed in this work.

An alternative way to deal with the second class constraints is to solve them explicitly.
Unfortunately, this can be done only breaking the explicit Lorentz covariance.
One way to do this is to use the parametrization \eqref{parameX}.
Then the kinetic term is diagonalized by the change of variables \eqref{solom} which gives
\be
\int\de^3 x\,\tP^a_{IJ}\p_t \omega_a^{IJ}=\int\de^3 x\Bigl[\tE_i^a\p_t \eta_a^{i}+\chi_i\p_t\omega^i\Bigr].
\ee
This shows that, whereas $(\eta_a^i,\tE^a_i)$ and $(\omega^i, \chi_i)$ are two canonical pairs, the variables $r_{ij}$
have vanishing conjugated momenta $p_r^{ij}=0$. These conditions play the role of additional primary constraints replacing
\eqref{conphi} of the covariant approach. To study their stability under time evolution, one needs to know the dependence
of the constraints \eqref{HPconstr} on $r_{ij}$. The expressions for all constraints in terms of the new variables can be found
in \cite{Alexandrov:1998cu}. The important fact is that only the Hamiltonian constraint $\PhiH$ carries a dependence on $r_{ij}$ so
that the stability condition generates the secondary constraint
\be
\frac{\p\PhiH}{\p r_{ij}}=0.
\ee
It is equivalent to \eqref{conpsi} and can be solved explicitly as \cite{Alexandrov:1998cu}
\be
\begin{split}
r_{ij}=&\, \frac{1}{1-\chi^2}\[
\(\hf\, \CX_{ij}\CX^{kl}-\CX_{i}^{\{ k}\CX_{j}^{l\}}\)\eps^{kmn}\tE^a_n\Et_b^{l}\p_a \tE^b_m
\right.
\\
&\, \left.
+\hf\(\delta_{ij}+\chi_i\chi_j\)\eps^{kmn}\tE^a_k\chi_m\p_a\chi_n
+\chi^{\{i}\eps^{j\}kl}\tE^a_k\p_a\chi_l
+\tE^a_{\{i}\eps_{j\}kl}\chi^k\eta_a^l\]
\end{split}
\label{solr}
\ee
with $\CX_{ij}=\delta_{ij}-\chi_i\chi_j$.
It can be checked that this solution, being substituted into \eqref{solom},
leads to the same commutation relations as \eqref{commomP}.

Using this non-covariant description, it is straightforward also to compute the following Poisson brackets
\be
\begin{split}
\{\omega_a^{IJ}(x), \tX^K(y)\} =&\, -\eta^{K[I} e_a^{J]}\delta(x,y),
\\
\{\omega_a^{IJ}(x), e_b^K(y)\} =&\, -2\(\Pt_b^{IJ}e_a^K+\Pt_b^{K[I} e_a^{J]}\)\delta(x,y).
\end{split}
\label{comm-om-eX}
\ee

\subsection{First class constraints}
\label{subsec-firstcl}

Defining the smeared version of the primary constraints
\be
\PhiG(n)=\int \de^3 x\, n^{IJ} \PhiG_{IJ},
\qquad
\PhiD(\vec N)=\int \de^3 x\, N^a \PhiD_a,
\qquad
\PhiH(\Nt)=\int \de^3 x\, \Nt \PhiH,
\ee
their algebra can be written in the following from
\be
\begin{split}
\{\PhiG(n) ,\PhiG(m)\}=&\,\PhiG(n\times m),
\\
\{\PhiD(\vec N) ,\PhiG(n)\}=&\,-\PhiG(N^a\p_a n),
\\
\{\PhiD(\vec N) ,\PhiG(\vec M)\}=&\,-\PhiG([\vec N,\vec M]),
\\
\{\PhiG(n),\PhiH(\Nt)\}=&\, 0,
\\
\{\PhiD(\vec N) ,\PhiH(\Nt)\}=&\,-\PhiH(\CL_{\vec N}\Nt),
\\
\{\PhiH(\Nt) ,\PhiH(\Mt)\}=&\, -\PhiD(\vec K_{N,M})+\PhiG(K_{N,M}^a\omega_a),
\end{split}
\ee
where
\be
\begin{split}
(n\times m)^{IJ}=&\, n^{IK}{m_K}^J-m^{IK}{n_K}^J,
\\
[\vec N ,\vec M ]^a=&\, N^b \partial_b M^a-M^b\p_b N^a,
\\
\CL_{\vec N}\Nt =&\, N^a\p_a \Nt-\Nt\p_a N^a,
\\
K_{N,M}^a=&\, g g^{ab}(\Nt\p_b\Mt-\Mt\p_b\Nt).
\end{split}
 \label{not1}
\ee
Since this algebra is closed, all these constraints are first class\footnote{In fact, $\PhiH$ does not weakly
commute with the secondary second class constraint. However, one can add to it a term proportional to the primary
second class constraint so that the full combination will be weakly commuting.}
and stable under time evolution.

For our calculations it is important to now how these constraints act on the fields $\tX^I$ and $e_a^I$.
This action can be computed using \eqref{comm-om-eX} and reads as
\be
\begin{split}
\{\PhiG(n),e_a^I\}=&\,-{n^I}_J e_a^J,
\\
\{\PhiG(n),\tX^I\}=&\,-{n^I}_J \tX^J,
\\
\{ \PhiD(\vec N),  e_a^I\} =&\, \p_a N^b e_b^I+N^b\p_b e_a^I,
\\
\{\PhiD(\vec N),  \tX^I\}=&\, \p_b(N^b \tX^I),
\\
\{\PhiH(\Nt), \tX^I\} =&\, g g^{ab} e_a^I \p_b \Nt-\Nt\(\PhiG^{IJ}\tX_J+2\tP^{a,IJ}D_a \tX_J\),
\\
\{\PhiH(\Nt), e_a^I\} =&\, \tX^I\p_a \Nt+\Nt\( D_a \tX^I+ \tX^I \Pt_a^{KL}\PhiG_{KL}\).
\end{split}
\label{constract}
\ee

\section{Useful relations}
\label{sec-useful}

The two fields, $\tP^a_{IJ}$ and $\Pt_a^{IJ}$, introduced in \eqref{canvar} and \eqref{defPt}, respectively,
and having the following decomposition
\beq
\tP^a_{IJ}&=&
\left\{
\begin{array}{ccl}
\hf\, \tE^a_i && \mbox{for } [IJ]=[0i],
\\
\tE^a_{[i}\chi_{j]}&\quad & \mbox{for } [IJ]=[ij],
\end{array}
\right.
\qquad
\Pt_a^{IJ}=
\left\{
\begin{array}{ccl}
\frac{\(\delta^i_j-\chi^i\chi_j\) \Et_a^j}{2(1-\chi^2)}&& \mbox{for } [IJ]=[0i],
\\
-\frac{\Et_a^i\chi^j-\Et_a^j\chi^i  \vphantom{\bigl)}}{2(1-\chi^2)}&\quad & \mbox{for } [IJ]=[ij],
\end{array}
\right.
\eeq
possess a set of nice properties. First of all, they are in a sense inverse of each other. Namely, they satisfy
\be
\tP^a_{IJ}\Pt_b^{IJ}=\hf\, \delta^a_b,
\qquad
\tP^{a}_{IJ}\Pt_a^{KL}=g^{-1}\tX_{[I}\delta_{J]}^{[K}\tX^{L]},
\ee
where the quantity on the r.h.s. of the last relation is a (half of) projector on bivectors collinear to the vector $\tX^I$.
Next, some of the contractions of these fields with the tetrad components are quite simple. In particular, one has
\be
\tP^a_{IJ} e_a^K= \delta_{[I}^K \tX_{J]},
\qquad
\tP^a_{IJ} e_b^J=-\hf\, \delta^a_b \tX_I,
\qquad
\Pt_a^{IJ} \tX_J=\hf\, e_a^I.
\ee
We also mention a few other relations which appear to be useful in calculations
\be
\begin{split}
g^{ab} e_a^I e_b^J-g^{-1} \tX^I \tX^J=&\,\eta^{IJ},
\\
\teps^{bcd}e_{b}^J e_{c}^K e_{d}^L = &\, -\eps^{JKLM} \tX_M,
\\
\teps^{bcd}{\eps^{IJ}}_{KL}\tX_J e_{c}^K e_{d}^L=&\, -2gg^{bc}e_c^I,
\\
\teps^{bcd}\eps_{IJKL}\tXp^I \eepi{a}^J \eemi{c}^K \eemi{d}^L=&\, 4\gp\Ptpi{a}^{IJ}\tPmi{IJ}^b,
\\
\teps^{bcd}\eps_{IJKL} \eepi{g}^I \eepi{f}^J \eemi{c}^K \eemi{d}^L=&\, -2\epst_{gfc}\eps^{IJKL}\tPpi{IJ}^c\tPmi{KL}^b,
\\
\frac14\, \mett_{ab}\teps^{bcd} {\eps^{IJ}}_{KL}\eepi{c}^K \eemi{d}^L=&\,\epst_{abc}\eps^{IJKL}\tPpi{KM}^b \tPmi{ML}^c,
\\
\eepi{a}^{[I}\tXm^{J]}=&\, \mett_{ab}\tP_{-}^{b,IJ}.
\end{split}
\label{usefulrel}
\ee
Finally, one can check the following formula for the inverse of the mixed metric
\be
\mett^{ab}=E_{-,i}^a E_{+,j}^b\(\delta^{ij}+\frac{\chip^i\chim^j}{1-\chip^k\chimi{k}}\)
=\frac{2\tPmi{IJ}^a\tP_{+}^{b,IJ}}{\eta_{KL}\tXp^K\tXm^L\vphantom{\tilde X}}.
\label{inversemet}
\ee
As $\mett_{ab}$ itself, it can be shown to be symmetric as a consequence of the constraints $\Scon^a$ \eqref{constrCcov}.


\end{document}